\documentclass[PRB,reprint,amsmath,amssymb,superscriptaddress,longbibliography,floatfix]{revtex4-1}

\usepackage[utf8]{inputenc}  
\usepackage[T1]{fontenc}     
\usepackage{lmodern}         
\usepackage[T1]{fontenc}
\usepackage{textcomp}
\usepackage{graphicx}

\usepackage{graphicx}        
\usepackage{dcolumn}         
\usepackage{bm}              
\usepackage{booktabs}        
\usepackage{multirow}        
\usepackage{amsmath} 
\usepackage{hyperref}
\hypersetup{
    colorlinks=true,        
    linkcolor=blue,         
    citecolor=blue,         
    urlcolor=blue,          
    breaklinks=true         
}

\usepackage{amssymb}
\usepackage{amsmath}

\usepackage{xcolor}

\begin{document}

\title{Enhanced anomalous Nernst effect in Pr-doped kagome and honeycomb magnet LaCo$_5$}

\author{Zheng Li}\thanks{}
\affiliation{School of Physics, Zhejiang University, Hangzhou 310058, China}

\author{Yueyang Wu}
\affiliation{School of Physics, Zhejiang University, Hangzhou 310058, China}

\author{Tianhao Li}
\affiliation{School of Physics, Zhejiang University, Hangzhou 310058, China}

\author{Jun-jian Mi}
\affiliation{School of Physics, Zhejiang University, Hangzhou 310058, China}

\author{Shu-Xiang Li}
\affiliation{School of Physics, Zhejiang University, Hangzhou 310058, China}

\author{Jiang Ma}
\affiliation{School of Physics, Zhejiang University, Hangzhou 310058, China}

\author{Qian Tao}
\affiliation{School of Physics, Zhejiang University, Hangzhou 310058, China}

\author{Xiaofeng Xu}
\affiliation{Key Laboratory of Quantum Precision Measurement of Zhejiang Province, School of Physics and Optical Engineering, Zhejiang University of Technology,Hangzhou, China}

\author{Sheng Xu}\thanks{}
\email{shengxu@zjut.edu.cn}
\affiliation{School of Physics, Zhejiang University, Hangzhou 310058, China}
\affiliation{Key Laboratory of Quantum Precision Measurement of Zhejiang Province, School of Physics and Optical Engineering, Zhejiang University of Technology,Hangzhou, China}

\author{Zhu-An Xu}
\email{zhuan@zju.edu.cn}
\affiliation{School of Physics, Zhejiang University, Hangzhou 310058, China}
\affiliation{State Key Laboratory of Silicon and Advanced Semiconductor Materials, Zhejiang University, Hangzhou, 310058, China}
\affiliation{Hefei National Laboratory, Hefei 230088, China}
\affiliation{Wuhan National High Magnetic Field Center, Huazhong University of Science \& Technology, Wuhan, 430074, China}

\date{\today}

\begin{abstract}
In this work, we report the successful synthesis of La$_{1-x}$Pr$_x$Co$_5$ ($x = 0.09, 0.21, 0.45$) single crystals. Magnetic field-dependent magnetization measurements reveal that Pr substitution induces negligible changes in the magnetic properties of LaCo$_5$, with the ferromagnetic ordering predominantly governed by the Co sublattices. Remarkably, the doped systems exhibit significant enhancements in the anomalous Nernst effect. At 300~K,  the anomalous Nernst thermopower $S^A_{yx}$ reaches 6.5~$\mathrm{\mu V/K}$ in La$_{0.55}$Pr$_{0.45}$Co$_5$, corresponding to a $\sim$ 40 \% enhancement compared to the parent compound. This significant improvement can be predominantly attributed to Pr-doping-induced Fermi level modification, which directly leads to a redistribution of Berry curvature across the Fermi surface. This work highlights the effectiveness of Pr doping in boosting the anomalous Nernst effect of La$_{1-x}$Pr$_x$Co$_5$, offering a practical strategy to design advanced materials for room-temperature energy-harvesting technologies and high-efficiency thermal sensing devices. 

\end{abstract}

\maketitle

The Nernst effect, a thermoelectric conversion mechanism that transforms thermal gradients into transverse electrical signals, holds significant promise for applications in waste heat recovery and highly sensitive thermal sensors \cite{1,2,3,4,5,6}. In conventional ferromagnetic systems, the anomalous Nernst thermopower ($S^A_{xy}$) empirically scales with the magnetization $S^A_{xy}=Q_S\mu_0M$, in which $Q_S$ is anomalous Nernst coefficient and ranges from 0.05 to 1~$\mathrm{\mu VK^{-1}T^{-1}}$ for trivial ferromagnets. Recent explorations of magnetic topological quantum materials have revealed that their non-zero Berry curvature can generate a giant anomalous Nernst effect (ANE) \cite{7,8,9,10,11,12,13,14}. 

The Nernst thermopower is expressed as \cite{15}, 
\begin{equation}
S_{xy}=-\frac{E_y}{\nabla_xT}=\frac{\sigma_{xx}\alpha_{xy}-\sigma_{xy}\alpha_{xx}}{\sigma_{xx}^2+\sigma_{xy}^2}
\end{equation}

\begin{equation}
\sigma_{xy}=-\frac{e^2}{\hbar}\int_{BZ}\frac{d^3k}{(2\pi)^3}f(k)\Omega_B(k)
\end{equation}

\begin{equation}
\alpha_{xy}=\frac{ek_B}{\hbar}\int_{BZ}\frac{d^3k}{(2\pi)^3}s(k)\Omega_B(k)
\end{equation}
where $\sigma_{xy}$ and $\alpha_{xy}$ are the Hall conductivity and Nernst conductivity, respectively, $e$ is the charge of an electron, $\hbar$ is the Planck’s constant $h$ divided by 2$\pi$, $k_B$ is the Boltzmann’s constant, $\Omega_B(k)$ is the Berry curvature, $f(k)$ is the Fermi-Dirac distribution function and $s(k)=-f(k)\mathrm{ln}(f(k))-(1-f(k))\mathrm{ln}(1-f(k))$. In magnetic topological materials, the $S^A_{xy}$ predominantly stems from the Berry curvature---an intrinsic property of the electronic band structure dictated by magnetic-ordering-induced time-reversal symmetry breaking and spin-orbit coupling---which demonstrates significant temperature dependence.\cite{15,16,17}. The $S^A_{xy}$ of Co$_3$Sn$_2$S$_2$ peaks at $\sim$2.9~$\mathrm{\mu V/K}$ at 80 K but diminishes sharply to $\sim$0.13~$\mathrm{\mu V/K}$ at 300 K \cite{9}. Similarly, Mn$_3$Sn exhibits a maximum  $S^A_{xy}$ of $\sim$0.6 $\mathrm{\mu V/K}$ at 160 K, which declines to $\sim$0.37 $\mathrm{\mu V/K}$ at 300 K \cite{10}. In contrast, TbMn$_6$Sn$_6$ demonstrates enhanced performance at elevated temperatures, achieving $S^A_{xy}$$\sim$ 2.15 $\mathrm{\mu V/K}$ at 300 K and further increasing to $\sim$2.4 $\mathrm{\mu V/K}$ at 320 K \cite{18}. Particularly noteworthy are Heusler alloys such as Co$_2$MnGa, which have emerged as leading candidates for large room-temperature ANE \cite{Co2MnGa,Co2MnGa2,Co2MnGa3,Co2MnGa4}. Xu \textit{et al.} reported a giant ANE thermopower of approximately $8~\mu$V/K at 300~K in Co$_2$MnGa single crystals \cite{Co2MnGa}.
Within the family of RCo$_5$ (R = Lanthanum rare earth ) compounds, which possess a crystal structure analogous to LaCo$_5$, an ANE thermopower of $3.1~\mu$V/K at 300~K was reported in SmCo$_5$ \cite{SmCo5}.
These examples highlight the ongoing efforts and diverse material platforms being explored for enhancing ANE, especially at room temperatures.

\begin{figure*}[t]
	\centering
	\includegraphics[width=0.7\textwidth]{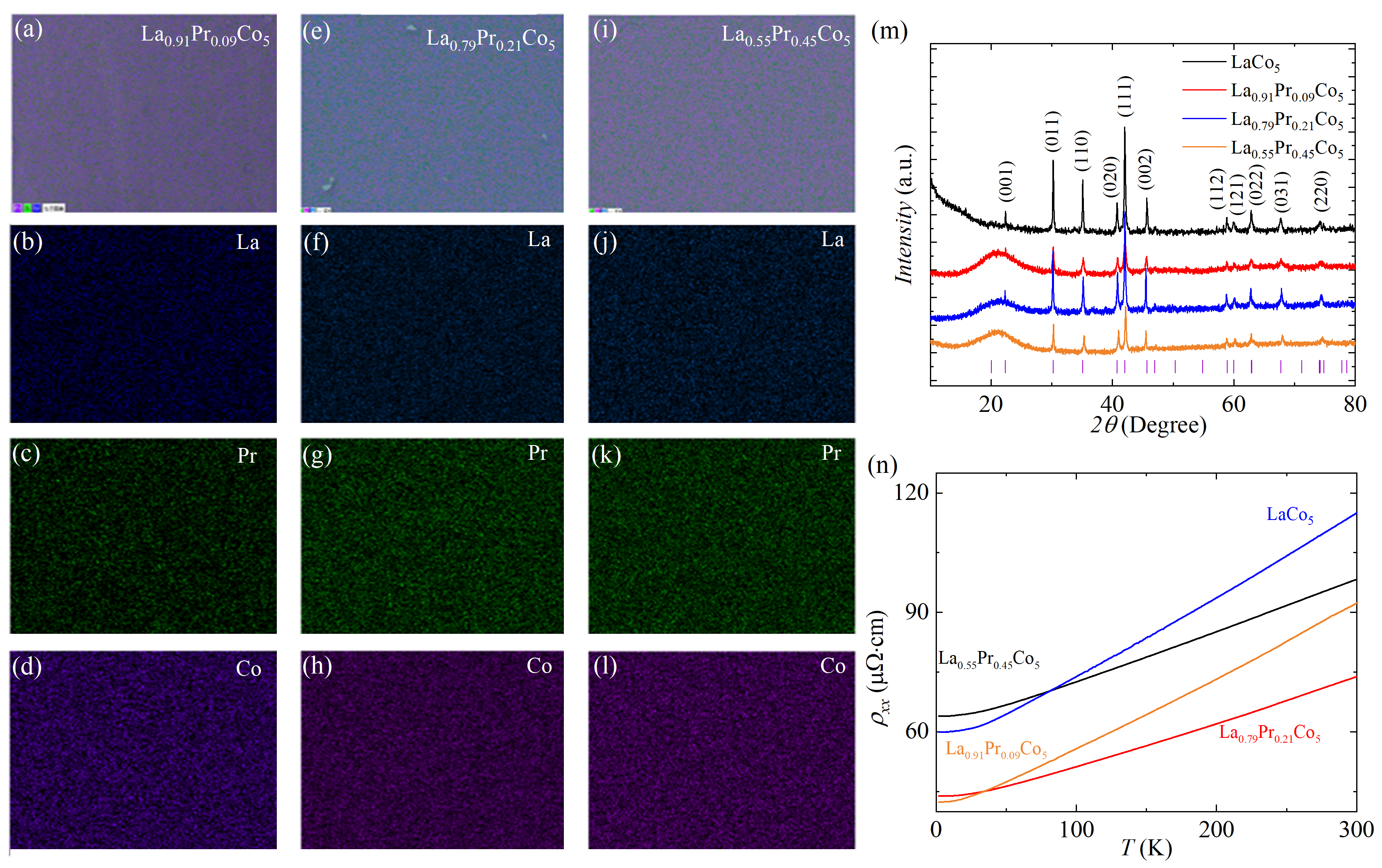}
	\caption{(a-d) SEM images with EDS elemental maps of La$_{0.91}$Pr$_{0.09}$Co$_5$, showing the spatial distributions of La, Pr, and Co. (e–h) SEM images with EDS elemental maps of La$_{0.79}$Pr$_{0.21}$Co$_5$, showing the spatial distributions of La, Pr, and Co. (i-l) SEM images with EDS elemental maps of La$_{0.55}$Pr$_{0.45}$Co$_5$, showing the spatial distributions of La, Pr, and Co. (m) Powder XRD patterns of the samples as-grown indexed to the $P6/mmm$ space group. (n) Temperature-dependent resistivity for La\({}_{1-x}\)Pr\({}_{x}\)Co\({}_{5}\). } 
	\label{fig1}
\end{figure*}%

\begin{figure}[htbp]
	\centering
	\includegraphics[width=0.48\textwidth]{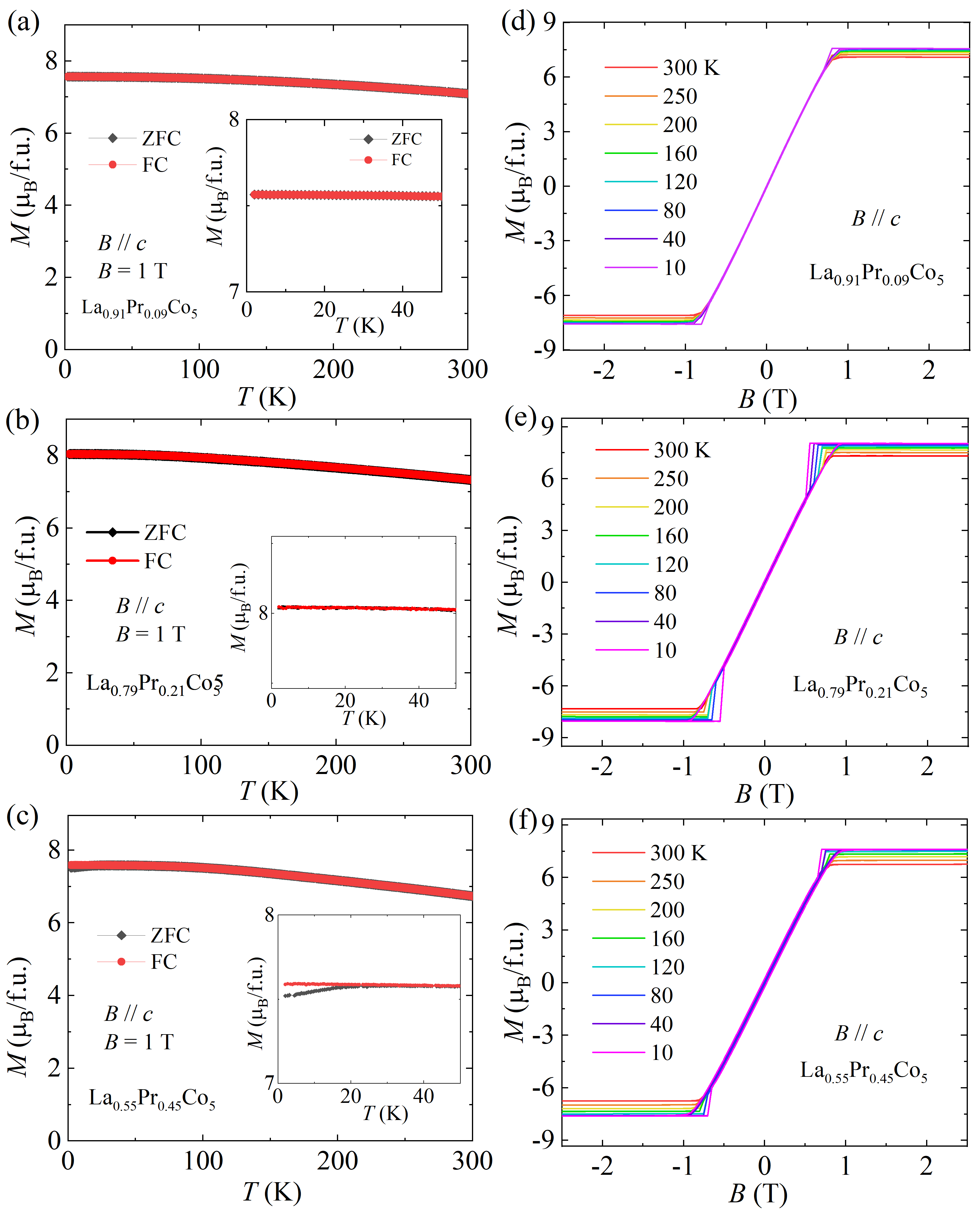}
	\caption{(a-c) Temperature-dependent magnetization for La\({}_{1-x}\)Pr\({}_{x}\)Co\({}_{5}\) with $B$=1 T applied along $c$-axis. (d-f) Field-dependent magnetization isotherms for La\({}_{1-x}\)Pr\({}_{x}\)Co\({}_{5}\). }
	\label{fig2}
\end{figure}%

In our prior work, we observed a large Berry curvature-driven room-temperature anomalous Nernst thermopower (4.6 $\mathrm{\mu V/K}$ at 300 K) in LaCo$_5$ single crystals---a kagome and honeycomb magnet with a high Curie temperature ($T_C$ = 840 K) \cite{21,22}. First-principles calculations revealed a dense distribution of Weyl points near the Fermi level, suggesting a strong Berry curvature contribution \cite{21}. To further optimize the ANE, in this work, we engineered Pr-doped LaCo$_5$ (La$_{1-x}$Pr$_x$Co$_5$, $x =0.09, 0.21$, 0.45) by introducing chemical pressure via partial Pr-for-La substitution. Crucially, this strategy preserved the ferromagnetic order, while manipulating the Fermi level to enhance Berry curvature effects. The optimized composition  La$_{0.55}$Pr$_{0.45}$Co$_5$ achieved $S^A_{yx}$ $\approx$ 6.5~$\mathrm{\mu V/K}$ at 300 K, representing a $\sim$40 \% enhancement over the parent compound.

\begin{figure*}[htb]
	\centering
	\includegraphics[width=0.8\textwidth]{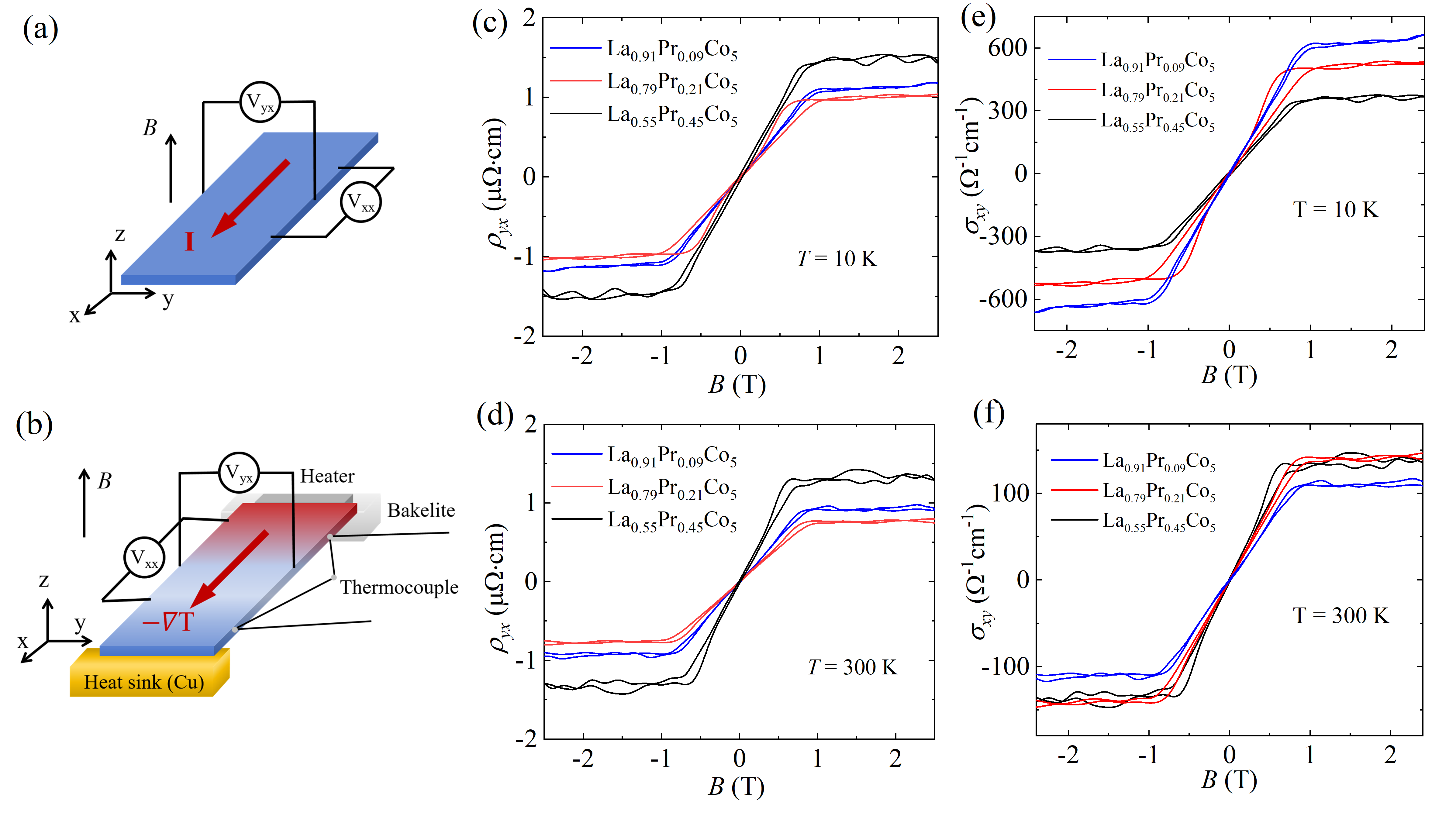}
	\caption{(a,b) Schematic illustrations for AHE and ANE measurements. (c,d) Field-dependent $\rho_{yx}$ for La\({}_{1-x}\)Pr\({}_{x}\)Co\({}_{5}\) with \textit{B} applied along \textit{c}-axis at 10 K and 300 K, respectively. (e,f) Field-dependent $\sigma_{xy}$ for La\({}_{1-x}\)Pr\({}_{x}\)Co\({}_{5}\) at 10 K and 300 K, respectively.} 
	\label{fig3}
\end{figure*}%

Single crystals of La$_{1-x}$Pr$_x$Co$_5$ were synthesized via the self-flux method. Stoichiometric mixtures of La/Pr ingots and Co powder with nominal ratios La : Pr : Co = 38 : 1.5 : 62 ($x=0.09$), 38 : 4 : 62 ($x=0.21$) and 38 : 10 : 62 ($x=0.45$) were sealed in alumina crucibles under vacuum within quartz tubes. The assemblies were heated to  1180 $\mathrm{^\circ C}$ followed by slow cooling to 920  $\mathrm{^\circ C}$ at 2 $\mathrm{^\circ C}$/h, with subsequent flux removal via centrifugation. The chemical homogeneity was verified through energy-dispersive spectroscopy (EDS) mapping, as shown in Figs. 1(a-l). The phase purity was confirmed by the powder X-ray Diffraction (XRD) patterns, as shown in Fig. 1(m). The refined lattice parameters are listed in Table I. The hexagonal lattice contracts along the $a$-axis  while it expands along the $c$-axis with increasing Pr content. Metallic behavior persists across both compositions, as displayed in  Fig. 1(n), evidenced by monotonically decreasing $\rho_{xx}(T)$ from 300 K to 2.5~K. Furthermore, the temperature-dependent magnetization with the magnetic field applied along the $c$-axis indicate that all three compounds remain ferromagnetic below 300 K, as shown in Figs. 2(a-c). The isothermal magnetization curves measured between 300 K and 10~K (Figs. 2(d-f)) demonstrate rapid saturation behavior, aligning with both the magnetic characteristics and saturation magnetization values of LaCo$_5$. This observation strongly suggests the persistence of robust ferromagnetic ordering at temperatures exceeding room temperature. The narrow hysteresis loops indicate weak soft ferromagnetism dominated by domain dynamics \cite{24}.

\begin{table}
	\centering
	\caption{The refined lattice parameters. }
	\label{oscillations}
	\renewcommand{\arraystretch}{1.5}
	\setlength{\tabcolsep}{3.3mm}{
		\begin{tabular}{cccccccc}
			
			\hline
			&                           &$a$ (\AA)    & $b$ (\AA)   & $c$ (\AA)&  \\
			\hline
			&LaCo$_5$                     &5.105      &5.105      &3.965   &\\
			&La$_{0.91}$Pr$_{0.09}$Co$_5$                     &5.103      &5.103      &3.978   &\\
			&La$_{0.79}$Pr$_{0.21}$Co$_5$                 &5.098     &5.098         & 3.983&\\
			
			&La$_{0.55}$Pr$_{0.45}$Co$_5$ &5.090      &5.090      &3.990  &\\
			\hline
		
	\end{tabular}}
\end{table}

\begin{figure*}[htb]
	\centering
	\includegraphics[width=0.9\textwidth]{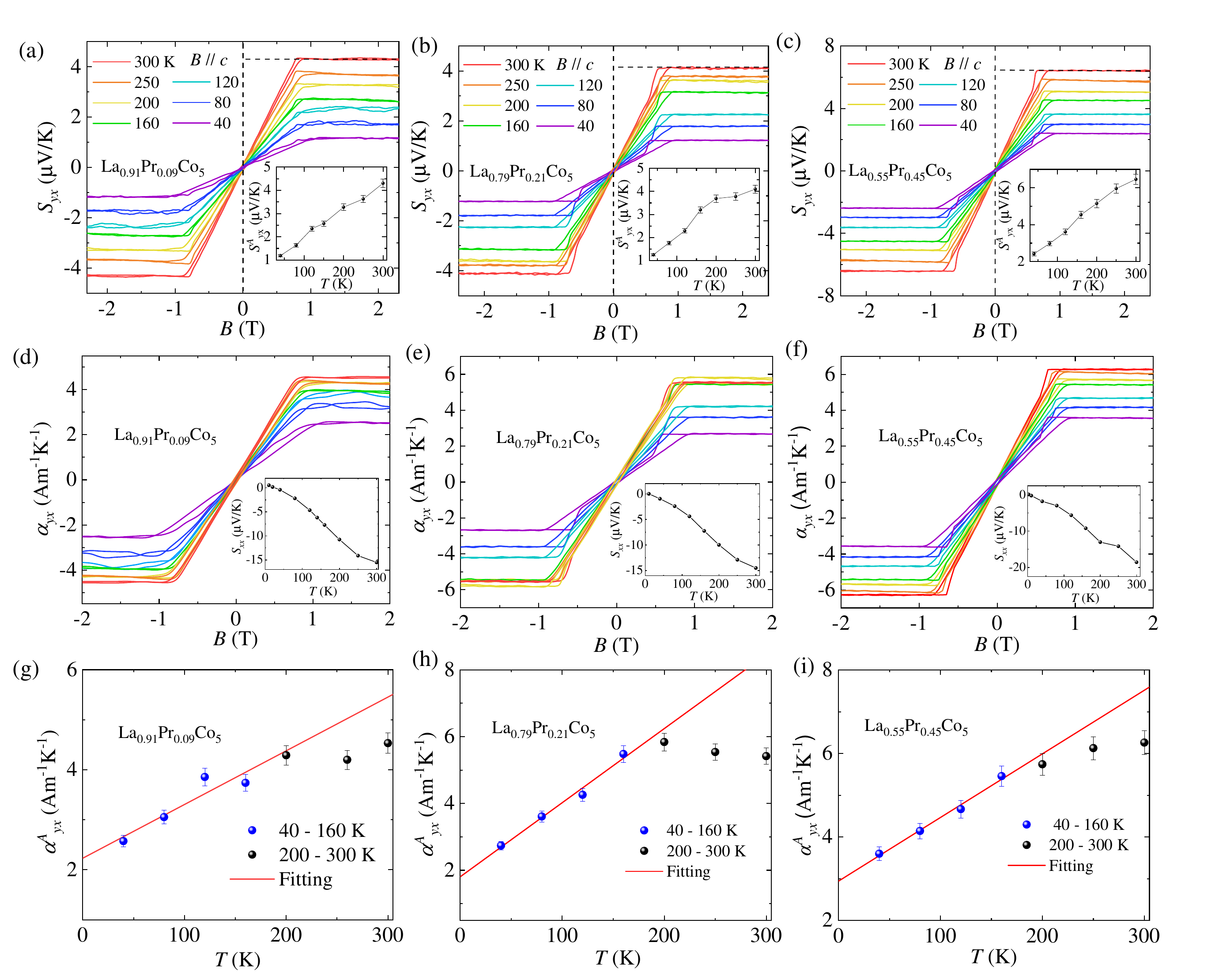}
	\caption{(a-c) Field-dependent $S_{yx}$  for La\({}_{1-x}\)Pr\({}_{x}\)Co\({}_{5}\) at different temperatures. (d-f) Field-dependent $\alpha_{yx}$ for La\({}_{1-x}\)Pr\({}_{x}\)Co\({}_{5}\). Insets: The corresponding temperature-dependent $S_{xx}$. (g-i) Temperature-dependent $\alpha^A_{yx}$ for La\({}_{1-x}\)Pr\({}_{x}\)Co\({}_{5}\).}
	\label{fig4}
\end{figure*}%

\begin{figure}[htb]
	\centering
	\includegraphics[width=0.47\textwidth]{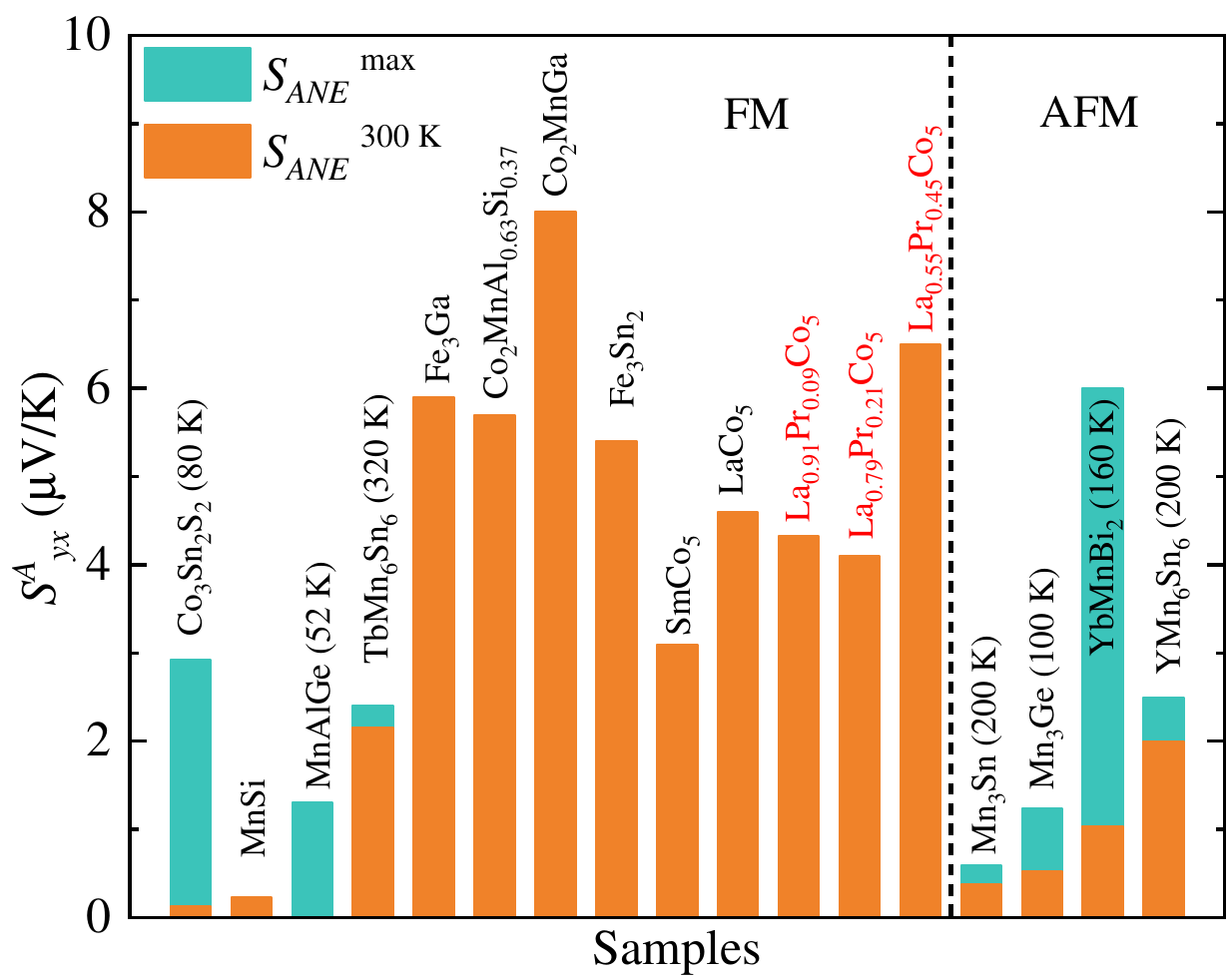}
	\caption{Comparison of anomalous Nernst thermopower between LaCo$_5$ system materials in the present study and previously reported ferromagnet, and antiferromagnet materials \cite{9,10,18,19,Co2MnGa,SmCo5,21,26,27,28,29,30,Co2MnAlSi}.
	}
	\label{fig5}
\end{figure}%

Figures~3(a, b) show schematic illustrations of the measurement configurations for the AHE and the ANE, respectively. The test samples were cut into a long strip shape with a thickness of approximately 0.1 mm.
To establish a thermal gradient for the thermoelectric measurements, one end of the sample was attached to a $1~\text{k}\Omega$ chip resistor (heater) and the other end to a heat sink.  This setup created a longitudinal temperature gradient ($-\nabla T$ $\sim$  0.4 - 1.0 K/mm) along the $x$-direction of the sample as shown in Fig.~3(b). Temperatures at two points along the sample length were monitored using thermocouples to determine the temperature gradient. The two junctions of this thermocouple were carefully placed in good thermal contact (via thermal grease, ensuring electrical insulation) with the side of the sample, aligned along its length and separated by a precisely measured distance. High-precision Keithley 2182A Nanovoltmeter was used to collect the voltage difference.  The Nernst and Hall signals were measured at both positive and negative magnetic field polarities, and the difference of the two polarities was deduced to remove the longitudinal contribution from the voltage probe misalignment.

The Pr-doped systems exhibit giant AHE characteristics, as shown in Figs. 3 (c,d). By linearly extrapolating the field-dependent $\rho_{yx}(B)$ data from the saturation regime to zero field, the $\rho^A_{yx}$ was obtained. For La$_{0.91}$Pr$_{0.09}$Co$_5$, the anomalous Hall resistivity ($\rho^A_{yx}$) reaches $\sim$1.15 $\mathrm{\mu\Omega\cdot cm}$  at 10 K, decreasing to $\sim$0.95 $\mathrm{\mu\Omega\cdot cm}$ at 300 K. In La$_{0.79}$Pr$_{0.21}$Co$_5$, the anomalous Hall resistivity ($\rho^A_{yx}$) reaches $\sim$1.00 $\mathrm{\mu\Omega\cdot cm}$  at 10 K, decreasing to $\sim$0.78 $\mathrm{\mu\Omega\cdot cm}$ at 300 K. Similarly, the $\rho^A_{yx}$ in La$_{0.55}$Pr$_{0.45}$Co$_5$ slightly decreases from $\rho^A_{yx}$ $\sim$1.51 $\mathrm{\mu\Omega\cdot cm}$ at 10 K to  $\sim$1.32 $\mathrm{\mu\Omega\cdot cm}$ at 300 K. The anomalous Hall conductivity (AHC), calculated via $\sigma^A_{xy} \approx \rho^A_{yx}/\rho_{xx}^2$, reveals critical doping effects: at 10 K, La$_{0.91}$Pr$_{0.09}$Co$_5$ achieves $\sigma^A_{xy}$ $\approx$ 635 $\mathrm{\Omega^{-1}cm^{-1}}$, exceeding both the La$_{0.55}$Pr$_{0.45}$Co$_5$ ($\sim$375 $\mathrm{\Omega^{-1}cm^{-1}}$) and La$_{0.79}$Pr$_{0.21}$Co$_5$ ($\sim$535 $\mathrm{\Omega^{-1}cm^{-1}}$). Room-temperature $\sigma^A_{xy}$ values converge to  $\sim$145 $\mathrm{\Omega^{-1}cm^{-1}}$ for both $x=0.21$ and $0.45$ systems---a 3 times enhancement over LaCo$_5$ ($\sim$50 $\mathrm{\Omega^{-1}cm^{-1}}$). In La$_{0.91}$Pr$_{0.09}$Co$_5$, the room-temperature AHC is about 110 $\mathrm{\Omega^{-1}cm^{-1}}$. Given the large number of Weyl points near the Fermi energy in the parent material LaCo$_5$, the variation in AHC could be attributed to the shift in the Fermi energy induced by Pr doping, which alters the Berry curvature integral over the entire Brillouin zone \cite{21}.

To further investigate the effect of Fermi level modification suggested by the AHC evolution under Pr doping, we conducted systematic ANE measurements---a critical probe of Berry curvature distribution near the Fermi surface \cite{15,20}, as the intrinsic ANE primarily originates from Berry curvature hotspots within thermally accessible states.  As shown in Figs. 4(a-c), the large ANE is observed in all of the three compounds, in which the $S^A_{yx}$ increase with the increasing temperature and reach a maximum of about 4.3 $\mathrm{\mu V/K}$  for $x=0.09$, 4.1 $\mathrm{\mu V/K}$  for $x=0.29$, and 6.5 $\mathrm{\mu V/K}$ for $x=0.45$, respectively. The value of $S^A_{yx}$ was determined by linearly extrapolating the field-dependent $S_{yx}(B)$ data from the saturation regime to zero field. To further study the origin of the ANE, the anomalous Nernst conductivity ($\alpha^A_{yx}$) is obtained by Eq. (1) and displayed in Figs. 4(d-f). At low temperature, the $\alpha^A_{yx}$ induced by the Berry curvature exhibits a linear behavior with temperature, $\alpha^A_{yx}=-\frac{\pi^2k^2_BT}{3\vert e\vert}(\frac{\partial\sigma_{yx}}{\partial E})_{E=E_F}$, following the Mott relation \cite{15,25}. As shown in Figs. 4(g-i), the $\alpha^A_{yx}$ of these three compounds exhibit a linear temperature dependence below 160 K, indicating the ANE in these three compounds stem from the intrinsic Berry curvature mechanism \cite{25-0}. Above 160 K, the $\alpha^A_{yx}$ deviates from the linear relationship similar to the parent  compound LaCo$_5$ \cite{21}. While the lattice parameters in this system evolve monotonically with increasing Pr doping concentration, the anomalous Nernst signal 
$S^A_{yx}$ exhibits non-monotonic behavior.  This discrepancy may originate from the dense distribution of Weyl nodes near the Fermi level \cite{21}, where even slight shifts in Fermi energy can lead to a nonlinear response in the Berry curvature. Such behavior is commonly observed in doped topological magnets, a representative example is found in In- and Ni-doped Co$_3$Sn$_2$S$_2$ films \cite{Co3Sn2S2}. 


The room-temperature ANE holds particular significance for practical applications. Figure 5 compares the $S^A_{yx}$ of LaCo$_5$-based materials with representative magnetic topological counterparts at 300 K, along with their respective peak-effect temperatures. Current benchmark kagome systems typically exhibit room-temperature $S^A_{yx}$ typically  below 6 $\mathrm{\mu V/K}$, whereas our optimized La$_{0.55}$Pr$_{0.45}$Co$_5$ achieves 6.5 $\mathrm{\mu V/K}$, which exhibits a $\sim$40 \% enhancement compared with the parent material LaCo$_5$. In addition, given the ultrahigh Curie temperature ($T_C$ = 840 K) of the parent LaCo$_5$ and the positive temperature coefficient of $S^A_{yx}$ in La$_{1-x}$Pr$_x$Co$_5$ ($x =0.09, 0.21, 0.45$), these systems exhibit promising potential for delivering enhanced $S^A_{yx}$ at elevated temperatures (400–600 K)—a critical regime for high-temperature waste heat recovery.

In summary, we synthesized single-crystalline La$_{1-x}$Pr$_x$Co$_5$ ($x = 0.09, 0.21, 0.45$) and demonstrated that Pr doping significantly enhances the ANE while preserving the robust ferromagnetic order. Remarkably, the optimized
La$_{0.55}$Pr$_{0.45}$Co$_5$ achieves a room-temperature anomalus Nernst thermopower 6.5 $\mathrm{\mu V/K}$, which is  40~\% improvement over LaCo$_5$ and suggested to stem from the Fermi level modification and Berry curvature redistribution. Crucially, the material maintains a high Curie temperature and exhibits a positive $S^A_{yx}$-$T$ correlation, suggesting further performance gains at elevated temperatures. This work establishes rare-earth doping as a scalable strategy. The synergy of high $T_C$, temperature-resilient ANE, and tunable Berry curvature opens avenues for room-temperature energy-harvesting technologies and high-efficiency thermal sensing devices.

\section*{Acknowledgments}
This work was supported by the National Natural Science Foundation of China (Grant No. 12574177; 12204410), the National Key R\&D Program of China (Grant No.2019YFA0308602), the Innovation program for Quantum Science and Technology (Grant No. 2021ZD0302500), the Zhejiang Provincial Natural Science Foundation of China (Grant No. LMS25A040002), and the Interdisciplinary program of Wuhan National High Magnetic Field Center (Grant No. WHMFC2025015), Huazhong University of Science and Technology

\section*{Author Declarations}
The authors have no conflicts to disclose.
\medskip
\section*{Data availability}
The data that support the findings of this study are available from the corresponding authors upon reasonable request.

\begin{thebibliography}{56}
\bibitem{1} X. Zhang and S.-C. Zhang, Proc. SPIE Int. Soc. Opt. Eng. \textbf{ 8373}, 837309 (2012).

\bibitem{2} K. Biswas, J. He, I. D. Blum, C.-I. Wu, T. P. Hogan, D. N. Seidman, V. P. Dravid, and M. G. Kanatzidis, Nature \textbf{489}, 414 (2012).

\bibitem{3} L.-D. Zhao, S.-H. Lo, Y. Zhang, H. Sun, G. Tan, C. Uher, C. Wolverton, V. P. Dravid, and M. G. Kanatzidis, Nature \textbf{508}, 373 (2014).

\bibitem{4} J. P. Heremans, R. J. Cava, and N. Samarth, Nat. Rev. Mater. 2, 17049 (2017).

\bibitem{5} J. He and T. M. Tritt, Science \textbf{357}, eaak9997 (2017).

\bibitem{6} J. Mao, H. Zhu, Z. Ding, Z. Liu, G. A. Gamage, G. Chen, and Z. Ren, Science \textbf{365}, 495 (2019).

\bibitem{7} H. Zhang, C. Q. Xu, and X. Ke, Phys. Rev. B \textbf{103}, L201101 (2021).

\bibitem{8} L. Ding, J. Koo, L. Xu, X. Li, X. Lu, L. Zhao, Q. Wang, Q. Yin, H. Lei, B. Yan, Z. Zhu, and K. Behnia, Phys. Rev. X \textbf{9}, 041061 (2019).

\bibitem{9} S. N. Guin, P. Vir, Y. Zhang, N. Kumar, S. J. Watzman, C. Fu, E. Liu, K. Manna, W. Schnelle, J. Gooth, C. Shekhar, Y. Sun, and C. Felser, Adv. Mater. 31, 1806622 (2019).

\bibitem{10} M. Ikhlas, T. Tomita, T. Koretsune, M.- T. Suzuki, D. Nishio-Hamane, R. Arita, Y. Otani, and S. Nakatsuji, Nat. Phys. \textbf{13}, 1085 (2017).

\bibitem{11} H. Zhang, J. Koo, C. Xu, M. Sretenovic, B. Yan, and X. Ke, Nat. Commun. 13, 1091 (2022). 

\bibitem{12} S. Roychowdhury, A. M. Ochs, S. N. Guin, K. Samanta, J. Noky, C. Shekhar, M. G. Vergniory, J. E. Goldberger and C. Felser, Adv. Mater. \textbf{34}, 2201350 (2022).

\bibitem{13} C. Wuttke, F. Caglieris, S. Sykora, F. Scaravaggi, A. B. Wolter, K. Manna, V. Suss, C. Shekhar, C. Felser, B. Buchner and C. Hess, Phys. Rev. B \textbf{100}, 085111 (2019).

\bibitem{14} A. Sakai, S. Minami, T. Koretsune, T. Chen, T. Higo, Y. Wang, T. Nomoto, M. Hirayama, S. Miwa, D. Nishio-Hamane, F. Ishii, R. Arita and S. Nakatsuji, Nature 581, 53–57 (2020).

\bibitem{15} D. Xiao, Y. Yao, Z. Fang, Q. Niu, Berry-phase effect in anomalous thermoelectric transport. Phys. Rev. Lett. \textbf{97}, 026603 (2006).

\bibitem{16}  D. F. Liu, A. J. Liang, E. K. Liu, Q. N. Xu, Y. W.Li, C. Chen, D. Pei, W. J. Shi, S. K. Mo, P. Dudin, T. Kim, C. Cacho, G. Li, Y. Sun, L. X. Yang, Z. K. Liu, S. S. P. Parkin, C. Felser, and Y. L. Chen, Science \textbf{365}, 1282–1285 (2019). 

\bibitem{17} E. Liu, Y. Sun, N. Kumar, L. Muechler, A. Sun, L. Jiao, S.-Y. Yang, D. Liu, A. Liang, Q. Xu, J. Kroder, V. Süß, H. Borrmann, C. Shekhar, Z. Wang, C. Xi, W. Wang, W. Schnelle, S. Wirth, Y. Chen, S. T. B. Goennenwein and C. Felser, Nat. Phys. \textbf{14}, 1125 (2018).
 
\bibitem{18} X. Xu, J.-X. Yin, W. Ma, H.-R. Tian, X.-B. Qiang, H. Zhou, J, Shen, H. Lu, T,-R. Chang, Z. Qu, S, Jia,  Nat. Commun. \textbf{13}, 1197 (2022). 

\bibitem{Co2MnGa} L. Xu, X. Li, L. Ding, T. Chen, A. Sakai, B. Fauque, S. Nakatsuji, Z. Zhu, K. Behnia,  Phys. Rev. B \textbf{101}, 180404(R) (2020). 

\bibitem{Co2MnGa2} A. Sakai, Y. P. Mizuta, A. A. Nugroho, R. Sihombing, T. Koretsune, M.-T. Suzuki, N. Takemori, R. Ishii, D. Nishio-Hamane, R. Arita, P. Goswami, S. Nakatsuji,  Nat. Phys. \textbf{14}, 1119-1124 (2018).

\bibitem{Co2MnGa3} S. N. Guin, K. Manna, J. Noky, S. J. Watzman, C. Fu, N. Kumar, W. Schnelle, C. Shekhar, Y. Sun, J. Gooth, C. Felser, NPG Asia Mater. \textbf{11}, 16 (2019).

\bibitem{Co2MnGa4} W. Zhou, A. Miura, T. Hirai, Y. Sakuraba, K.-i. Uchida. Appl. Phys. Lett. \textbf{122}, 062402 (2019).

\bibitem{SmCo5} A. Miura, H. Sepehri-Amin, K. Masuda, H. Tsuchiura, Y. Miura, R. Iguchi 
, Y. Sakuraba, J. Shiomi, K. Hono, K. Uchida,  Appl. Phys. Lett. \textbf{115}, 222403 (2019).



\bibitem{19} S. N. Guin, K. Manna, J. Noky, S. J. Watzman, C. Fu, N. Kumar, W. Schnelle, C. Shekhar, Y. Sun, J. Gooth, C. Felser, NPG Asia Mater. \textbf{11}, 16 (2019).

\bibitem{20} A. Sakai, Y.P. Mizuta, A.A. Nugroho, R. Sihombing, T. Koretsune, M.-T. Suzuki, N. Takemori, R. Ishii, D.
Nishio-Hamane, R. Arita, P. Goswami, S. Nakatsuji, Nat. Phys. \textbf{14}, 1119 (2018).

\bibitem{21} S. Xu, L. Zhou, S.-X. Li, X.-Y. Zeng, C. Jiang, J.-J. Mi, Z. Li, T.-L. Xia, and Z.-A. Xu, Mater. Today. Phys. \textbf{38}, 101269 (2023). 

\bibitem{22} W.A.J.J. Velge, K.H.J. Buschow, J. Appl. Phys. \textbf{39}, 1717 (1968).

\bibitem{24} Z. Yang, M. Lange, A. Volodin, R. Szymczak, and V. V. Moshchalkov, Nat. Mater. \textbf{3}, 793 (2004).



\bibitem{25} M. Cutler, N. F. Mott, Observation of Anderson localization in an electron gas. Phys. Rev. \textbf{181}, 1336–1340 (1969).

\bibitem{25-0} N. Nagaosa, J. Sinova, S. Onoda, A.H. MacDonald, N.P. Ong, Rev. Mod. Phys. \textbf{82}, 1539 (2010).

\bibitem{Co3Sn2S2} S. Noguchi, K. Fujiwara, Y. Yanagi, M. Suzuki, T. Hirai, T.
Seki, K. Uchida, and A. Tsukazaki, Nat. Phys. \textbf{20}, 254 (2024).


\bibitem{mott} Solid State Physics. Von N. W. Ashcroft und N. D. Mermin; Holt, Rinehart and Winston, New York , XXII, 826 Seiten, (1976).

\bibitem{26} Y. Hirokane, Y. Tomioka, Y. Imai, A. Maeda, Y. Onose, Phys. Rev. B \textbf{93}, 014436 (2016).




\bibitem{27} S. N. Guin, Q. Xu, N. Kumar, H. Kung, S. Dufresne, C. Le, P. Vir, M. Michiardi, T. Pedersen, S. Gorovikov, S. Zhdanovich, K. Manna, G. Auffermann, W. Schnelle, J. Gooth, C. Shekhar, A. Damascelli, Y. Sun, C. Felser, Adv. Mater. \textbf{33}, 2006301 (2021).

\bibitem{28} Z. Fan, S. Minami, S. Akamatsu, A. Sakai, T. Chen, D. Nishio-Hamane, and S. Nakatsuji, Adv. Funct. Mater. \textbf{32}, 2206519 (2022).

\bibitem{29} Y. Li, J. Zhou, M. Li, L. Qiao, C. Jiang, Q. Chen, Y. Li, Q. Tao, Z.-A. Xu, Phys. Rev. Appl. \textbf{19}, 014026 (2023).

\bibitem{30} Y. Pan, C. Le, B. He, S.J. Watzman, M. Yao, J. Gooth, J.P. Heremans, Y. Sun, C. Felser, Nat. Mater. \textbf{21}, 203–209 (2020).

\bibitem{Co2MnAlSi} Y. Sakuraba, K. Hyodo, A. Sakuma, and S. Mitani, Phys. Rev. B \textbf{101}, 134407  (2020).


\end{thebibliography}

\end{document}